\documentclass[10pt,conference]{IEEEtran}

\pdfoutput=1

\usepackage{cite}
\usepackage{amsmath,amssymb,amsfonts}
\usepackage{graphicx}
\usepackage{textcomp}
\usepackage{xcolor}
\usepackage[hyphens]{url}
\usepackage{booktabs}
\usepackage{tabularx}
\usepackage{multirow}
\usepackage{bm}
\usepackage{tikz}
\usepackage[caption=false,font=footnotesize]{subfig}
\usepackage[linesnumbered,ruled,vlined]{algorithm2e}
\usepackage{pifont}
\usepackage{afterpage}
\usepackage[hidelinks]{hyperref}

\definecolor{myred}{HTML}{EA6B66}

\newcommand{\xname}{PLoRA}
\newcommand{\cmark}{{\color{green!55!black}\ding{51}}}
\newcommand{\xmark}{{\color{red!75!black}\ding{55}}}

\newcommand{\circlednumber}[3][white]{%
  \tikz[baseline=(char.base)]{
    \node[
      shape=circle,
      fill=#2,
      text=#1,
      draw=#1,
      inner sep=0.6pt
    ] (char) {#3};
  }%
}

\title{\xname{}: An NDP-Enhanced Pooled-Memory\\ System for Cost-Efficient Multi-LoRA Serving}

\IEEEoverridecommandlockouts

\newcommand\hpcaauthors{%
  Zhongkai Yu\textsuperscript{1,*}, Ohm Rishabh Venkatachalam\textsuperscript{1,*},
  Zheng Wang\textsuperscript{1}, Yikai Li\textsuperscript{2},
  Yichen Lin\textsuperscript{1}, Zihao Yu\textsuperscript{3},
  Yuke Wang\textsuperscript{4},\\
  Liu Liu\textsuperscript{5}, Xulong Tang\textsuperscript{6},
  Shuyi Pei\textsuperscript{7}, Yangwook Kang\textsuperscript{7},
  Yufei Ding\textsuperscript{1}}

\newcommand\hpcaaffiliation{%
  \textsuperscript{1}University of California, San Diego \quad
  \textsuperscript{2}New York University \quad
  \textsuperscript{3}University of California, Santa Cruz \quad
  \textsuperscript{4}Rice University\\
  \textsuperscript{5}Rensselaer Polytechnic Institute \quad
  \textsuperscript{6}University of Pittsburgh \quad
  \textsuperscript{7}Samsung Semiconductor, Inc.}

\newcommand\hpcaemail{%
  \{zhy055, ovenkatachalam, zhw100, yil384, yufeiding\}@ucsd.edu \quad
  yl9321@nyu.edu \quad zyu165@ucsc.edu\\
  yuke.wang@rice.edu \quad liu.liu@rpi.edu \quad tax6@pitt.edu \quad
  \{shuyi.pei, yangwook.k\}@samsung.com}

\author{%
  \IEEEauthorblockN{\hpcaauthors{}%
    \thanks{\textsuperscript{*} Contributed equally to this work.}}
  \IEEEauthorblockA{%
    \hpcaaffiliation{} \\
    \hpcaemail{}
  }
}

\begin{document}
\maketitle
\thispagestyle{plain}
\pagestyle{plain}

\bstctlcite{IEEEexample:BSTcontrol}

\begin{abstract}
Multi-LoRA serving is how one base model becomes thousands of specialized
variants, one adapter per user, task, or agent, and the deployments
can hold 1000-plus adapters. Serving them is hard
because the workload inverts what GPUs provide: terabytes of memory against
only tens of TFLOPS, and because every published system stages its adapters
from CPU DRAM over PCIe, where each access pays a kernel stop and a host-run
copy and capacity ends at the motherboard's DIMM slots. Meanwhile,
memory-semantic fabrics such as CXL and NVLink are converging on pooled memory
that an accelerator addresses with its own loads and stores, and near-data
processing (NDP) can place compute beside the pooled data. How to serve
multi-LoRA workloads on such hardware remains unexplored.

This paper introduces PLoRA, an NDP-enhanced pooled-memory system for
cost-efficient multi-LoRA serving. PLoRA keeps adapters and KV cache in the
pool and returns only reduced results over the link, through a read-compute
interface the GPU drives with its own loads and stores. Above this
architecture, a GPU memory management system picks among four LoRA and two
attention execution strategies for each adapter and caches the most
performance-critical bytes in GPU memory, guided by a link-parameterized cost
model. On one H100 serving 1000 adapters, PLoRA attains the lowest decode
latency on every model and workload we measure, averaging 6.6$\times$ below a
real-machine S-LoRA at under 3.4\% added device area. The link itself stops
mattering: throughput saturates at 32\,GB/s on short contexts, a quarter of CXL 3.1, and the
verdict survives scale: per-GPU demand falls from 7B to a modeled 1.2T
deployment once adapter traffic shards with the tensor parallelism. The
design runs unchanged from CXL-class to NVLink-class fabrics, and surplus
bandwidth buys pooled capacity rather than speed.
\end{abstract}

\section{Introduction}
\label{sec: intro}
Every large language model now ships as a family. Fine-tuning specializes one base model into variants for programming assistance~\cite{llm_programming1, llm_programming2}, translation~\cite{gpt_llm_trans1, lu2024llm_trans2}, and chat~\cite{llm_chatbot1, llm_chatbot2}, and low-rank adaptation (LoRA) has made that specialization cheap: each variant is a small matrix pair, an adapter, trained over frozen base weights~\cite{lora, qlora, fine_tune_adapter, fine_tune_prefix}.
Deployments therefore serve thousands of adapters at once, one per user for personalization, one per task for accuracy, and one per agent in multi-agent frameworks.
Serving systems share the base model, S-LoRA and Punica batch all adapters over one copy of the weights~\cite{sheng2023s-slora, chen2024punica}, but the adapters themselves overflow the GPU: 1000 adapters over a 13B model at 4K context need over 1.3\,TB~\cite{llama2, llama3}.
Every published system stages that overflow from host DRAM over PCIe~\cite{sheng2023s-slora, chen2024punica, wu2024dlora, li_caraserve_2024}.
Across production and research alike, adapter capacity lives behind the host.

\begin{table}[tb]
    \centering
    \caption{Memory and computation consumption}
    \label{table:1_mem_compute}
    \footnotesize
    \begin{tabular}{@{}cccc@{}}
    \toprule
                         & A100      & H100      & Multi-LoRA Infer \\ \midrule
    Mem Space               & 80 GB     & 80 GB     & 1320 GB              \\
    Computation          & 312 TFLOPS & 990 TFLOPS & 28 TFLOPS              \\
    Mem / Compute & 0.26      & 0.08      & 47.14                \\ \bottomrule
    \end{tabular}
    \vspace{-10pt}
\end{table}

The workload has outgrown that arrangement.
Multi-LoRA inference inverts the ratio a GPU is built for, asking two to three orders of magnitude more bytes per FLOP than the hardware supplies (\autoref{table:1_mem_compute}): at the 800 tokens/s Punica reports~\cite{chen2024punica}, it consumes only 28 TFLOPS of FP16 compute against the roughly 1000 TFLOPS a modern GPU provides~\cite{ampere_white_paper, hopper_white_paper}, while demanding more than 15 times one GPU's memory, so buying capacity in units of GPUs buys mostly idle silicon.
And the host path that avoids this cost pays in kind.
PCIe caps transfers an order of magnitude below local GPU DRAM~\cite{pcie_bw1, pcie_bw2}, the accelerator cannot address host memory, so every access pays a kernel stop and a CPU-initiated copy~\cite{pcie_latency}, and host capacity ends at the motherboard's DIMM slots~\cite{motherboard_limit}.

Meanwhile, the interconnect industry is converging on exactly the missing ingredient.
CXL 3.1 switches pool memory that an accelerator addresses with its own load and store instructions~\cite{cxl3.1}.
NVLink-C2C already carries the CXL protocol between NVIDIA parts at several times CXL's bandwidth, NVLink Fusion is opening that fabric to third-party silicon~\cite{nvlink_fusion, nvlink_fusion_press}, and UALink standardizes the same load/store semantics for accelerator-attached memory~\cite{ualink}.
Memory-semantic pooled memory is arriving, whether or not serving systems are ready to use it.

This convergence invites a rethinking of where adapter capacity belongs: why not serve multi-LoRA workloads directly out of the pool, with compute beside the data, so that the link stops mattering?
\textbf{Our insight} is that link bandwidth is the scarce resource, near-data compute can be spent in its place, and for this workload the exchange rate is extremely favorable.
Two measurements set that rate.
First, the memory-hungry part of multi-LoRA serving is also the compute-light part: KV cache and adapter operations take 97.8\% of the memory but only 3.0\% of the computation (\autoref{table:1_breakdown}).
Second, those operations reduce heavily. The attention product $\bm{P}_{1\times N} = \bm{Q}_{1\times D}\bm{K}^{T}_{D\times N}$ emits a result $D \geq 4096$ times smaller than the operands it consumes.
Moving operands across the link therefore costs thousands of times more than moving results, and a small compute agent beside the pooled memory converts the one into the other.
The multiplier is set by the operator's reduction ratio, not by the fabric, so a little near-data silicon stands in for a great deal of interconnect.

\textbf{The design is therefore a property of the link's semantics, not of any one standard.}
\xname{} needs exactly three things from the interconnect. \textbf{F1:} the accelerator reaches pooled memory with its own load/store instructions, with no host in the data path. \textbf{F2:} a device-side compute agent is reachable through that same interface. \textbf{F3:} capacity scales independently of host DIMM slots.
Compute Express Link (CXL) 3.1 provides all three and is the only openly specified option with shipping silicon today~\cite{cxl3.1}, so we build and evaluate that instantiation, and we express the cost model over link bandwidth and latency rather than over CXL.

\begin{figure}[t]
    \centering
    \includegraphics[width=\columnwidth]{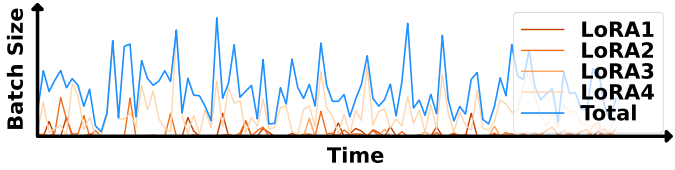}
    \caption{Batch sizes differ across adapters, and each adapter's batch varies over time.}
    \label{fig:2-multi-lora}
    \vspace{-6pt}
\end{figure}

Pooling with near-data compute is necessary but not sufficient, because two challenges remain.
\textbf{(1) Strategy coordination.} With compute on both sides of the link, each adapter can be served in several ways, and the best way depends on its instantaneous batch size, which varies across adapters and over time (\autoref{fig:2-multi-lora}). Neither the strategies nor the online mechanism that picks among them is explored.
\textbf{(2) Resource contention.} Adapters and their KV caches compete for GPU memory, so per-adapter optima are rarely globally feasible, and the system needs a global allocator that spends scarce GPU memory on the most performance-critical bytes.

We present \xname{}, an NDP-enhanced pooled-memory system that serves multi-LoRA workloads cost-efficiently on one accelerator and, to our knowledge, the first to serve them on memory-semantic pooled memory with near-data processing.
Across an interconnect ladder, \xname{} stops responding to link bandwidth at 32\,GB/s on short contexts, 4$\times$ below the CXL 3.1 link we assume and 28$\times$ below NVLink 5, where a decode step takes exactly as long as streaming the base model from HBM, and by 64\,GB/s on long.
Near-data execution does not merely relieve the link, it removes the link from the critical path and leaves pooled memory performing as though it were local.
The verdict survives scale: from 7B to a modeled 1.2T tensor-parallel deployment, per-GPU bandwidth demand falls as models grow, provided adapter traffic shards with the model.
We hope this work gives pooled memory a first-class role in LLM serving, and tells memory-semantic fabrics where silicon is best spent: capacity rather than speed.
We contribute:
\begin{itemize}
    \item A device architecture that adds NDP cores to memory-semantic expanders, plus a read-compute interface the accelerator drives with load/store semantics and no host mediation (\autoref{sec: architecture}).
    \item Four GPU/NDP strategies for LoRA and two for attention, spanning the trade-off between GPU memory consumed and bytes crossing the link (\autoref{sec: collaborative}).
    \item A link-parameterized cost model and an online algorithm that choose a strategy per adapter under a GPU memory budget (\autoref{system}).
    \item An interconnect-ladder evaluation showing that near-data execution substitutes for link bandwidth from 7B to a modeled 1.2T deployment, porting the design across CXL-class and NVLink-class links (\autoref{sec: link-generality}).
\end{itemize}

\section{Background}

\subsection{LLM and LoRA}
\noindent\textbf{Large language models} serve a request in two stages~\cite{attention, pd}. Prefill consumes the input and emits the first response token while building a Key-Value (KV) cache for the input tokens. Decode then emits one token per iteration and appends to that cache, which grows until it requires terabytes, so \xname{} offloads it to CXL for cost-effective serving.

\noindent\textbf{Low-rank adaptation} freezes the base weights and attaches small trainable matrices, the LoRA adapters, to the query, key, and value generation, to the projection, and to the FFN layers~\cite{lora}.
For a base weight $\bm{W}\in\mathbb{R}^{D_{1}\times D_{2}}$, LoRA adds $\bm{A}\in\mathbb{R}^{D_{1}\times R}$ and $\bm{B}\in\mathbb{R}^{R\times D_{2}}$ with rank $R \ll \min(D_1, D_2)$, so the layer computes $\bm{y} = \bm{x}(\bm{W}+\bm{AB})$.

\subsection{Multi-LoRA Serving}
\label{subsec: multi-lora-serving}
Multi-LoRA serving deploys over 1000 adapters against one shared base model, which is efficient because adapter parameters are hundreds of times smaller than the base~\cite{fine_tune_peft}.

Two diversity factors complicate it: as~\autoref{fig:2-multi-lora} shows, one adapter's batch swings over time and batches differ across adapters, which makes strategy selection hard (\autoref{subsec:why-difficult}).

S-LoRA~\cite{sheng2023s-slora} and Punica~\cite{chen2024punica} decouple base model and adapter inference, and dLoRA~\cite{wu2024dlora} switches between separate and integrated execution based on request distribution. All three offload adapters to system DRAM and pay for the movement back. Replacing that DRAM with pooled memory and NDP capability (\autoref{subs:CXL}) opens the design space that this paper explores.

\begin{table}[t]
\centering
\caption{Interconnects that can carry pooled memory. Single-number NVLink rates are bidirectional.}
\label{tab:fabrics}
\footnotesize
\setlength{\tabcolsep}{2.5pt}
\begin{tabular}{@{}lcccccc@{}}
\toprule
Interconnect & Per dir. & Aggr. & F1 & F2 & F3 & Open \\
 & (GB/s) & (GB/s) & & & & spec \\ \midrule
PCIe 5.0 $\times$16 (host offload) & 64 & 128 & \xmark & \xmark & \xmark & \cmark \\
CXL 2.0 $\times$8 expander~\cite{micron_cz120} & 32 & 64 & \cmark & \cmark & \cmark & \cmark \\
CXL 3.x $\times$16~\cite{cxl3.1} & 128 & 256 & \cmark & \cmark & \cmark & \cmark \\
UALink 1.0~\cite{ualink} & \multicolumn{2}{c}{200 GT/s per lane} & \cmark & \cmark & \cmark & \cmark \\
NVLink-C2C, GH200~\cite{gracehopper_indepth} & 450 & 900 & \cmark & \cmark & \cmark & \xmark \\
NVLink 5, Blackwell~\cite{nvlink_fusion} & 900 & 1800 & \cmark & \cmark & \cmark & \xmark \\
NVLink 6, Rubin~\cite{nvlink_scaleup} & 1800 & 3600 & \cmark & \cmark & \cmark & \xmark \\ \bottomrule
\end{tabular}
\vspace{-6pt}
\end{table}

\subsection{Memory-Semantic Pooled Memory}
\label{subs:CXL}
\xname{} assumes a pool of memory that the accelerator addresses with its own load and store instructions. We call such an interconnect \emph{memory-semantic}, and we depend on it only through three properties: (F1) the accelerator issues loads and stores to the pool with no host in the data path, (F2) a device-side compute agent is reachable through that same interface, and (F3) pooled capacity scales independently of host DIMM slots.
\autoref{tab:fabrics} surveys the fabrics that provide these properties.

\noindent\textbf{CXL} is an open standard built on the PCIe physical layer that supplies F1 through F3 today~\cite{cxl3.1}.
It defines a mandatory CXL.io sub-protocol for basic transfers plus optional CXL.cache and CXL.memory for coherence and memory access, and classifies devices by which they implement. \textbf{Type 2} devices, processors with attached memory such as GPUs, implement all three, while \textbf{Type 3} devices are memory expanders implementing CXL.io and CXL.memory.
\xname{} builds on Type 3 devices, adding compute units restricted to local data reduction, which distinguishes them from Type 2 devices whose compute units address the whole pool.
CXL 3.1 peer-to-peer support lets an accelerator reach a Type 3 device directly, and switching scales a pool to 4096 devices.

\noindent\textbf{Proprietary scale-up fabrics} supply the same three properties at higher bandwidth.
Fourth-generation NVLink accesses peer memory ``using direct loads, stores, and atomic operations''~\cite{gracehopper_indepth}, Extended GPU Memory already lets GPU threads address a CPU-attached LPDDR5X pool over NVLink-C2C~\cite{cuda_egm}, and NVIDIA has patented memory-only endpoints, fabric-attached memory, hanging directly off the NVLink fabric~\cite{nvidia_fam_patent1, nvidia_fam_patent2}, so a near-data expander on this class of link is anticipated rather than hypothetical.
NVLink Fusion now licenses the interface to third-party CPUs and accelerators~\cite{nvlink_fusion, nvlink_fusion_press, arm_nvlink, sifive_nvlink}, and NVLink-C2C carries the CXL protocol itself~\cite{nvidia_nvlink_c2c}, the same protocol on a wider wire.
The opening is partial, a Fusion deployment still needs one NVIDIA part and the management layer stays closed~\cite{nvlink_proprietary, nextplatform_nvlink}, and no NVLink-attached device today is built primarily for memory expansion~\cite{werner2025nvlinkc2c}.
UALink 1.0 pursues the same semantics openly for up to 1024 accelerators~\cite{ualink}, and NVIDIA has paid over \$900 million to license a memory-fabric technology that pools 18\,TB of CXL DDR5~\cite{nvidia_enfabrica, enfabrica_emfasys}, which prices what a pooled-memory path is worth to a GPU vendor.
Our claim is conditional and deliberately so: if a memory-semantic scale-up link admits a memory expander, \xname{} runs on it unchanged, and \autoref{sec: link-generality} shows headroom for every fabric listed.

Every entry in \autoref{tab:fabrics} except PCIe host offload satisfies all three, yet they span an order of magnitude in bandwidth, so we parameterize the cost model by link bandwidth and latency (\autoref{subsec: cost model}) and sweep that range in \autoref{sec: link-generality}.

\section{Motivation}
\subsection{Why is CXL + NDP a Possible Solution?}
\begin{table}[t]
    \centering
    \caption{Mem and compute breakdown of a 13B model}
    \label{table:1_breakdown}
    \footnotesize
    \begin{tabular}{@{}cccc@{}}
    \toprule
             & KV Cache Ops & LoRA Ops & Model Weight Ops \\ \midrule
    Memory  & 76.3\%  & 21.5\%    & 2.3\%   \\
    Compute  & 2.4\%  & 0.6\%     & 97.0\%  \\ \bottomrule
    \end{tabular}
\vspace{-6pt}
\end{table}
\noindent\textbf{CXL memory extension}. We decompose inference into KV cache operations, LoRA adapter computations ($\bm{Y} = \bm{XAB}$), and base model computations.
In~\autoref{table:1_breakdown} (batch=200, seq\_length=2048) the first two consume 97.8\% of memory but only 3\% of computation, so they belong on the CXL device.

\noindent\textbf{NDP cores}. CXL supplies capacity but limited bandwidth, so NDP cores pay off on high-reduction operations: $\bm{P}_{1\times N}=\bm{Q}_{1\times D}\times \bm{K}^{T}_{D\times N}$ reduces its input by $D \ge 4096$, so computing it in the device sends only the small result over the link.

\subsection{Why is CXL+NDP better than PCIe+CPU?}
\label{subsec: why is CXL better?}

\begin{figure}[t]
    \centering
    \includegraphics[width=0.40\textwidth]{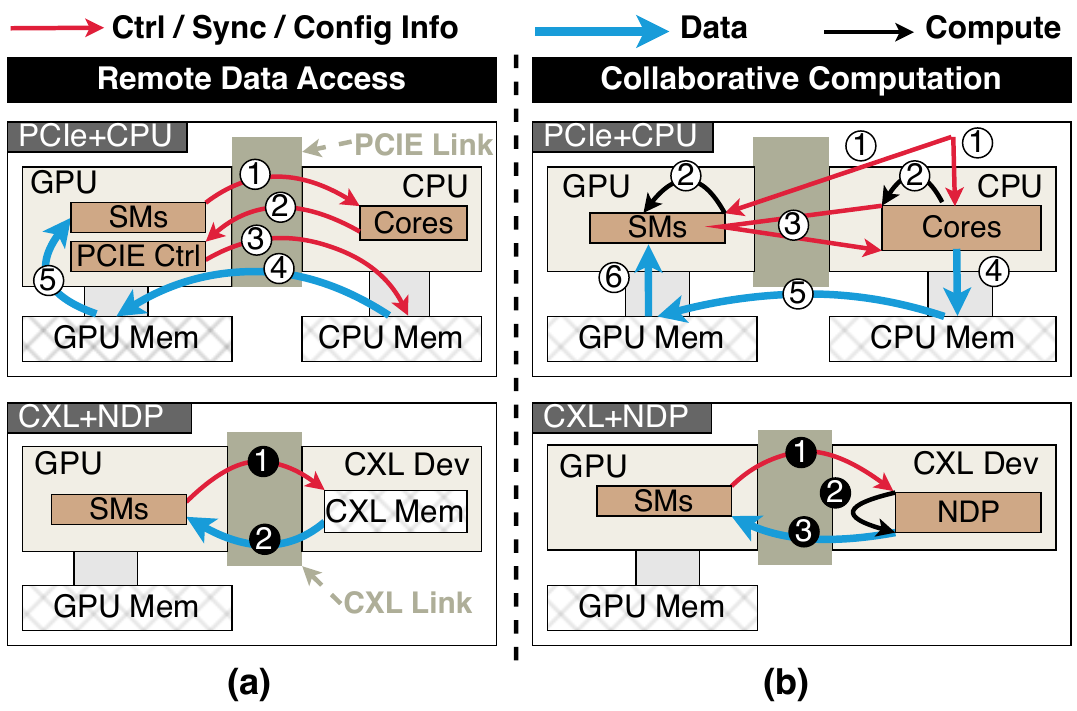}
    \caption{Pooled memory with NDP against the PCIe+CPU offload path, in (a) remote data access and (b) collaborative computation, with CXL as the example fabric.}
    \label{fig:3-compare}
\vspace{-8pt}
\end{figure}

\textbf{(1) Low remote memory access overhead.}
In~\autoref{fig:3-compare}(a), host-staged data reaches the SMs only across a kernel boundary: the running kernel finishes and returns control to the CPU, the CPU runs the copy program, a \texttt{cudaMemcpy} that stages the data over PCIe into GPU memory, and the next kernel reads it, so a kernel stop and a host-driven copy serialize ahead of the first byte. Our CPU-LoRA-Offload and PLoRA-NoCXL baselines quantify this path.
With CXL the SMs load and store directly, in two steps.

\textbf{(2) Efficient collaborative computation.}
Splitting work with the CPU is worse, as \autoref{fig:3-compare}(b) shows, because the two launch separate programs, synchronize over PCIe, and copy results back, which repeats the path of item (1) on every exchange.
\xname{} instead has the GPU issue commands the device runs locally.

\textbf{(3) Capacity and aggregate bandwidth that scale.} Host DRAM tops out near 1\,TB of DIMM capacity at a few hundred GB/s, short of the 1.3\,TB the workload already needs, whereas a switched CXL pool grows its aggregate internal bandwidth with every device added. One GPU still reads that data over a fixed link, so \xname{} spends the aggregate near the data.

\begin{figure}[t]
    \centering
    \subfloat[]{\includegraphics[width=0.49\linewidth]{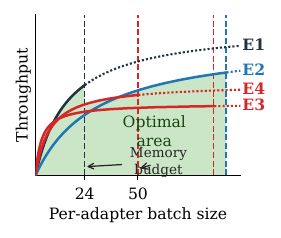}\label{fig:3-exe-strategy}}
    \hfill
    \subfloat[]{\includegraphics[width=0.49\linewidth]{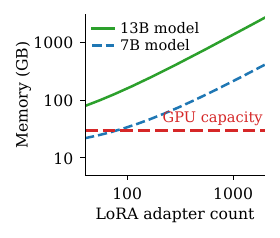}\label{fig:3-gpu mem}}
    \caption{(a) The best of the four strategies for one LoRA adapter changes with its batch size. (b) Memory demand exceeds GPU capacity once every adapter takes its own optimum.}
    \vspace{-10pt}
\end{figure}

\begin{figure}[t]
    \centering
    \includegraphics[width=0.44\textwidth]{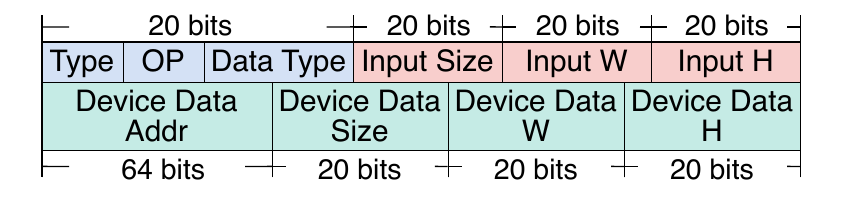}
    \caption{The read-compute request descriptor.}
    \label{fig:4-inst}
\vspace{-8pt}
\end{figure}

\begin{figure*}[t!]
    \centering
    \includegraphics[width=0.95\textwidth]{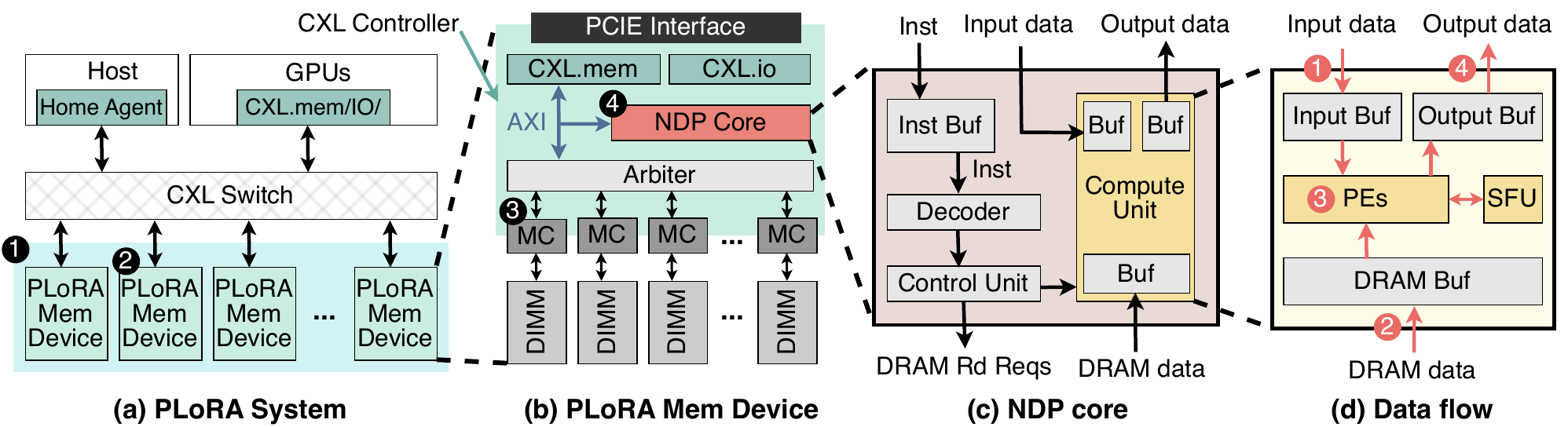}
    \caption{The \xname{} system and architecture.}
    \label{fig:4-architecture}
\vspace{-7pt}
\end{figure*}

\subsection{Why is it Difficult to Serve Multi-LoRA Inference even with CXL + NDP?}
\label{subsec:why-difficult}
\textbf{1. The optimal execution strategy of each LoRA adapter varies.}
Compute and memory sit on both the GPU and the device, so every adapter admits several execution strategies. In~\autoref{fig:3-exe-strategy}, the black line E1 is GPU-only execution without CXL memory, the blue line E2 offloads to CXL memory but leaves the NDP cores idle, and the red lines E3 and E4 are our collaborative strategies (\autoref{subsec: lora adapter computation}). Dotted segments give theoretical performance and only solid segments are achievable under the GPU memory constraint. The best strategy is highly sensitive to each adapter's batch size, which varies in both ways of~\autoref{subsec: multi-lora-serving}.

\textbf{2. LoRA adapters and their associated KV cache compete for GPU memory}.
In~\autoref{fig:3-gpu mem}, a few adapters fit, but as the adapter count grows the demand overshoots GPU capacity, sometimes by more than 10$\times$, which rules out the per-adapter optimum. Strategy selection, KV cache placement, and the GPU memory budget form one decision space that request arrivals keep shifting, and \autoref{system} handles them together.

\section{\xname~Architecture}
\label{sec: architecture}

\subsection{System and Architecture Overview}
\label{subsec: architecture overview}
The \xname{} system in \autoref{fig:4-architecture}(a) has a host CPU, GPUs, and \xname{} memory devices behind a CXL switch.
Each GPU is a CXL type-2 device with 10-80 GB of DRAM (HBM) and every \xname{} memory device ({\small \circlednumber{black}{2}}) a CXL type-3 device, so together they form a memory pool ({\small \circlednumber{black}{1}}) expanding GPU memory to tens of TBs.
CXL 3.1 peer-to-peer communication lets the GPU reach this pool directly, so each GPU carries CXL.mem and CXL.io IP.
We leave the GPU's CXL.cache agent unused for the \xname{} region, a per-region choice the specification permits without loss of conformance, because coherence adds little to LLM workloads with regular large-granularity accesses and GPU cache line state may not match the CPU-centric protocols of CXL.cache.

In \autoref{fig:4-architecture}(b), each \xname{} memory device pairs a CXL controller with DRAM channels, and an AXI Network on Chip (NoC) joins the controller's CXL.mem IP, CXL.io IP, an arbiter with Memory Controllers (MC), and our NDP core ({\small \circlednumber{black}{3}}).
Aggregated DRAM bandwidth exceeds 1 TB/s~\cite{park2024lpddr_cxl}, far above the 128 GB/s per direction of the CXL 3.1 link, so NDP cores spend internal bandwidth to relieve that bottleneck.
Vendors already list CXL devices with FPGA on their roadmaps~\cite{cmmh, cmmh_char}, so the integration is practical.
GPU memory therefore holds the compute-intensive base weights, while the large but compute-light adapters and KV cache live on the devices and are cached on the GPU only when capacity allows.

\subsection{Near Data Processing Core Design}
\label{subsec: NDP}
The NDP cores assist two computations whose inputs already sit on the devices, LoRA adapter computation and KV cache attention.
In \autoref{fig:4-architecture}(c), the instruction buffer holds instructions until they issue, and the decoder turns each into control code for the control unit, which drives the DRAM controller and the Compute Unit.
Previous CXL+NDP work~\cite{zhou2024neomem_tiering, jang2023cxl_anns, park2024lpddr_cxl, recxl, ham2024lowoverhead_pnm4} covers the lower-level details we omit.
The Compute Unit holds an input buffer for GPU data, a DRAM buffer for device-resident data, and an output buffer for intermediate and final results, and its logic pairs PEs optimized for GEMV and GEMM with a Special Function Unit (SFU) for softmax.
We follow Samsung's HPCA'24 CXL-PNM device~\cite{park2024lpddr_cxl}, with 1.1\,TB/s bandwidth, 512\,GB capacity, and 4.9 TFLOPS, while our NDP cores use 2 TFLOPS, so the configuration is silicon-backed, not optimistic.

\subsection{Read Compute Request Design}
\label{subsec: readcompute}
A read-compute request extends the load and store interface with a command directing the NDP core to compute on GPU-supplied operands plus device-resident data and return only the result.
\autoref{fig:4-architecture}(d) traces the flow from GPU input ({\small \circlednumber{myred}{1}}) and device-resident data ({\small \circlednumber{myred}{2}}) through the PEs and SFU ({\small \circlednumber{myred}{3}}) to the result returned to the GPU ({\small \circlednumber{myred}{4}}).
As \autoref{fig:4-inst} shows, a request packs Type (load, store, read-compute), OP (GEMM, GEMV, Softmax) and data type into 20 bits, the 64-bit base address of the device-resident operand, and the byte size and two dimensions of that operand and of the GPU-supplied input at 20 bits each. This 204-bit layout needs no parsing, fits one 256-Byte CXL 3.1 flit, and supplies the address the base and limit registers below check.

\textbf{Issuing and protocol.} Read-compute is a memory-mapped command interface, not a new CXL link transaction, with CXL.io MMIO for control and standard CXL.mem access to a Host-managed Device Memory (HDM) region for data. The GPU writes the operand and descriptor into an HDM submission queue over CXL.mem, writes a per-descriptor valid flag last, then rings a CXL.io MMIO doorbell. CXL.io and CXL.mem are not mutually ordered, so the doorbell is only a wakeup hint and the in-HDM valid flag is the ordering point gating the descriptor pull into the instruction buffer of \autoref{fig:4-architecture}(c). A reserved HDM window and a device-side address filter separate commands from ordinary accesses by address, adding no CXL.mem opcode and staying CXL 3.1 compliant.

\textbf{Coherence and completion.} The instruction buffer is private device SRAM the GPU never addresses. With CXL.cache disabled the shared HDM region has no hardware cache coherence, reducing consistency to a coarse-grained producer-consumer handoff with no back-invalidation. The device writes the result and then a completion flag to HDM, which the GPU polls or takes as an MSI-X interrupt. Per-requester queues bound to a PASID isolate concurrent GPUs, and per-queue base and limit registers check every device-side address.

\textbf{Portability to other fabrics.} Nothing in the above is specific to CXL, which matters because CXL is the openly specified carrier rather than the fastest one (\autoref{subs:CXL}).
It needs only a device memory window the accelerator can write with ordinary stores, a doorbell register, and an ordering point the device can observe, which is why it introduces no new link opcode.
On any fabric exposing device memory as loadable and storable accelerator address space, the submission queue, the valid flag, and the completion flag land in that window unchanged, and the doorbell becomes a control register write instead of a CXL.io MMIO write.
Word-granularity load, store, and atomic access to a memory-only fabric endpoint is the model NVIDIA already describes for fabric-attached memory~\cite{nvidia_fam_patent1, nvidia_fam_patent2} and ships through Extended GPU Memory~\cite{cuda_egm}, and NVLink-C2C carries the CXL protocol itself~\cite{nvidia_nvlink_c2c}, so the queue, the flags, and the doorbell keep their encodings and only the wire changes.
A natively coherent fabric would make our software handoff redundant and a mutually ordered one would remove the valid-flag ordering point, so the read-compute interface is a lower bound on such fabrics, and \autoref{sec: link-generality} prices the extra bandwidth.

\section{Collaboration between GPU and NDP Cores}
\label{sec: collaborative}

\begin{figure}[tb]
    \centering
    \includegraphics[width=0.48\textwidth]{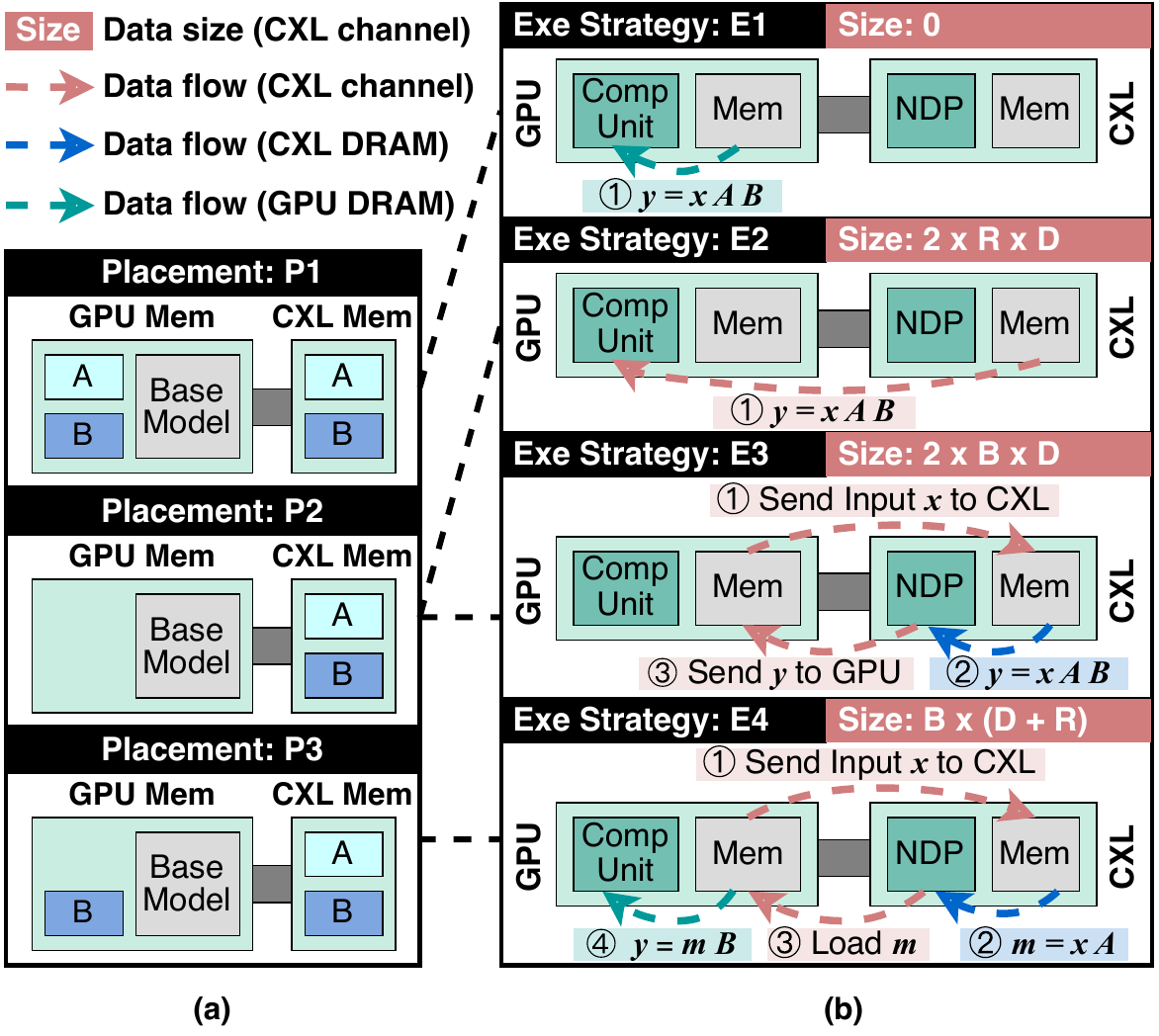}
    \caption{Three data placements with varying GPU memory demand enable four execution strategies.}
    \label{fig:5-placement-exe}
\vspace{-7pt}
\end{figure}

\subsection{LoRA Adapter Computation}
\label{subsec: lora adapter computation}
GPU compute units directly access both local GPU DRAM (HBM) and remote \xname{} memory devices, so no data movement is mandatory and GPU memory instead serves as a software-managed cache for the devices.

As \autoref{fig:5-placement-exe}(a) shows, each adapter consists of two matrices, $\bm{A}$ and $\bm{B}$, initially stored in \xname{} memory devices due to the GPU memory constraint.
\xname{} caches adapters in GPU memory to accelerate $\bm{y} = \bm{x}\bm{A}\bm{B}$ in three ways that suit different GPU memory budgets: caching both $\bm{A}$ and $\bm{B}$ (P1), caching neither (P2), or caching only $\bm{B}$ (P3).

\autoref{fig:5-placement-exe}(b) shows the four matching execution strategies.
Under P1 the GPU computes $\bm{y}$ from the cached $\bm{A}$ and $\bm{B}$ and loads nothing from the devices (E1).
Under P2, where no data resides in GPU memory, two strategies are possible: the GPU compute units calculate $\bm{y}$ directly from $\bm{A}$ and $\bm{B}$ in the devices (E2), or a device NDP core computes $\bm{y} = \bm{x}\bm{A}\bm{B}$ and sends $\bm{y}$ to the GPU (E3).
Under P3 the NDP core computes the intermediate result $\bm{m} = \bm{x}\bm{A}$ and transfers $\bm{m}$ to the GPU for the subsequent computation $\bm{y} = \bm{m}\bm{B}$ (E4).

No strategy is universally optimal, as each trades GPU memory consumption against the transfer volume that \autoref{fig:5-placement-exe}(b) lists, and that volume determines performance because link bandwidth bottlenecks the entire \xname{} system.
Three parameters set it: the per-adapter batch size $B_{i}$, 0 to 256, the LoRA dimension $R$, 2 to 128, and the model dimension $D$, 4096 to 16384, with $B_{i}$ the most dynamic since requests arrive and depart constantly.
Large $B_{i}$ favors E1 and E2, whose transfer volume is independent of it, and small $B_{i}$ favors E3 and E4, whose transfer volume scales with it.
A strategy selection algorithm in \autoref{subsec: strategy selection algorithm} therefore picks the best strategy for each adapter from the current system state.

\begin{figure}[tb]
    \centering
    \includegraphics[width=0.46\textwidth]{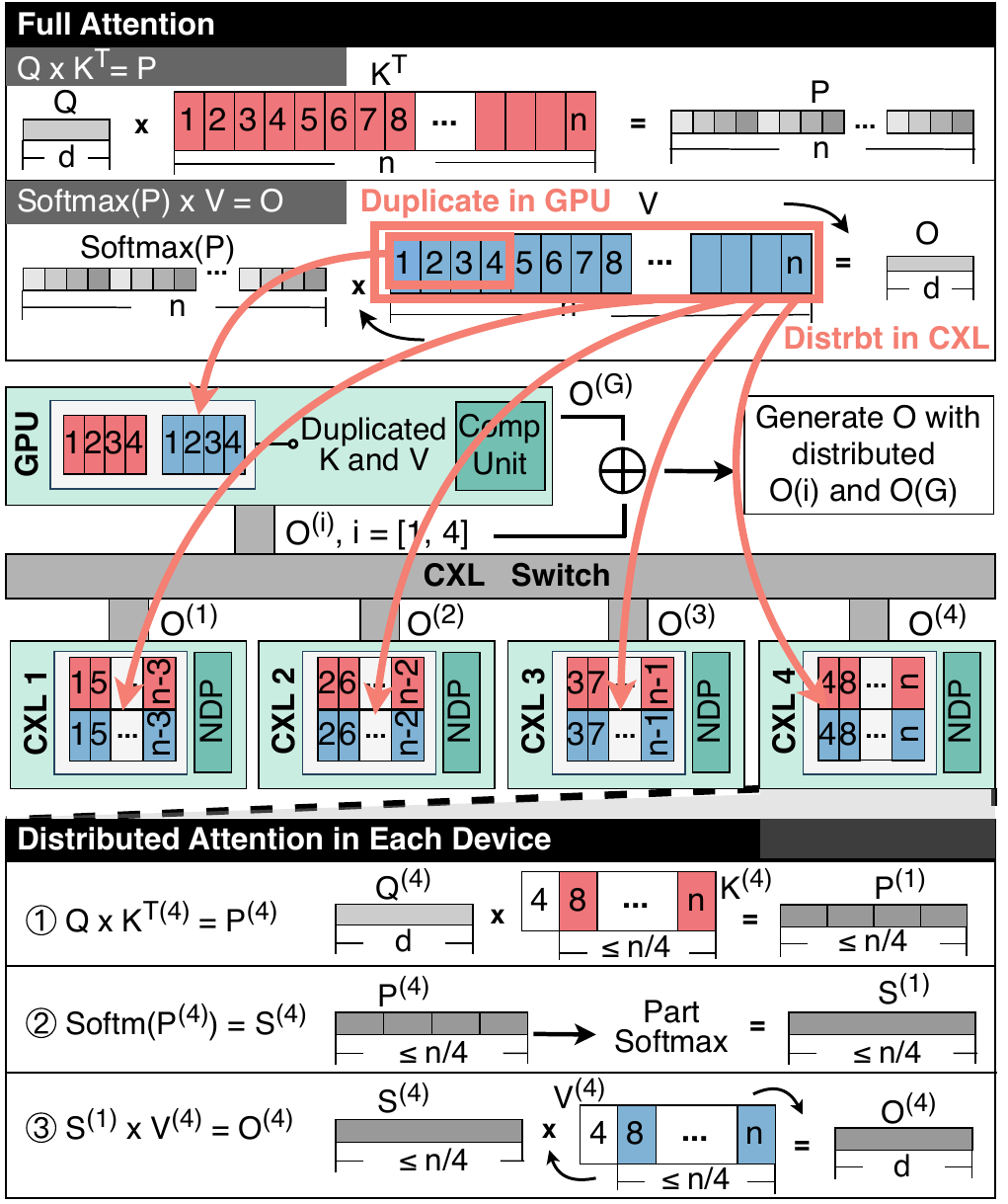}
    \caption{A collaborative approach for attention computation utilizing GPU and all \xname{} memory devices.}
    \label{fig:5-distributed-attention}
\end{figure}

\subsection{Attention Computation}
\label{subsec: attention computation}
The attention computation involves three steps, \textcircled{1} $\bm{P} = \bm{Q} \bm{K}^{T}$, \textcircled{2} $\bm{S} = softmax(\bm{P})$, and \textcircled{3} $\bm{O} = \bm{S} \times \bm{V}$, and it raises two challenges that two strategies address.
First, Steps \textcircled{1} and \textcircled{3} are memory-bounded GEMV operations, so slow KV cache migration from the devices to the GPU limits their efficiency.
Second, offloading part of Steps \textcircled{1} and \textcircled{3} to NDP cores still sends the intermediate result $\bm{S}$ over the link, which adds significant latency at long sequence lengths.

\textbf{1. KV Cache Distribution Strategy}: For collaborative GEMV in Steps \textcircled{1} and \textcircled{3}, \xname{} evenly distributes the KV cache of each request across all memory devices at the token level, storing the KV cache of the $i^{th}$ token in the $(i \bmod N_{\text{dev}})^{th}$ device, as shown in~\autoref{fig:5-distributed-attention}.
\xname{} also duplicates a small portion of the KV cache in GPU memory (HBM) when space permits, which lets GPU and NDP cores compute in parallel and maximizes aggregated internal bandwidth.
\xname{} overlaps prefill KV cache transfer with computation, and during decode each new token's KV cache goes directly to the devices.
Because duplicated entries compete with LoRA adapters for GPU memory, the GPU management system in~\autoref{system} evicts them using our algorithm and cost model.

\textbf{2. Integral Execution Strategy}: To keep the intermediate result $\bm{S}$ off the link, each device independently completes all three steps on its own share of the partitioned KV cache.
However, Softmax in Step 2 needs all input elements to produce each output element, which rules out a naive distribution.
Each device therefore applies the partial Softmax technique of FlashAttention~\cite{flashattention1, flashattention2, flashattention3}, computing Softmax over partial data and retaining an anchor value for integrating the overall output.
The \xname{} memory devices do not communicate with each other. The GPU gathers its own partial output together with the per-device partial outputs and anchors over CXL.mem and rescales them into the exact global output, so the partial results may return in any order.

\section{\xname{} GPU Memory Management System}
\label{system}

When a request arrives, \xname{} looks up its LoRA adapter ID and the current batch size, then runs the strategy selection algorithm (\autoref{subsec: strategy selection algorithm}) and the cost model (\autoref{subsec: cost model}) to determine the optimal execution strategy for all requests that share this adapter.
A new request can, for example, push an adapter's batch past the point where caching both matrices beats fetching them, switching that adapter from E4 to E1, loading its A matrix, and evicting KV cache to fit the budget.

\begin{table*}[t]
\centering
\caption{Per-strategy cost terms for LoRA computation.}
\label{tab:cost-terms}
\setlength{\tabcolsep}{3pt}
\resizebox{\ifdim\width>\textwidth \textwidth\else \width\fi}{!}{%
\footnotesize
\begin{tabular}{@{}cccc@{}}
\toprule
 & $T_{\text{GPU\_Ej}}$ & $T_{\text{dev\_cpt\_Ej}}$ & $T_{\text{dev\_trs\_Ej}}$ \\ \midrule
E1 & $\max\{\frac{4 \Sigma_{i=1}^{N} B_{i1}\times D \times R_{i}}{C_{\text{GPU}}},
\frac{2 \times D \times S \times \Sigma_{i:B_{i1}>0} R_{i}}{W_{\text{GPU}}}\}$ & $0$ & $0$ \\[2pt]
E2 & $\max\{\frac{4 \Sigma_{i=1}^{N} B_{i2}\times D \times R_{i}}{C_{\text{GPU}}},
2L_{\text{link}} + \frac{2 \times D \times S \times \Sigma_{i:B_{i2}>0} R_{i}}{W_{\text{link}}}\}$ & $0$ & $0$ \\[2pt]
E3 & $0$ & $\max\{\frac{4\Sigma_{i=1}^{N} B_{i3}\times D \times R_{i}}{C_{\text{dev}}\times N_{\text{dev}}},
\frac{2 \times D \times S \times \Sigma_{i:B_{i3}>0} R_{i}}{W_{\text{DRAM}}\times N_{\text{dev}}}\}$
& $2L_{\text{link}} + \frac{2D \times S\times\Sigma_{i=1}^{N} B_{i3} }{W_{\text{link}}}$ \\[2pt]
E4 & $\max\{\frac{2\Sigma_{i=1}^{N} B_{i4}\times D \times R_{i}}{C_{\text{GPU}}},
\frac{D \times S \times \Sigma_{i:B_{i4}>0} R_{i}}{W_{\text{GPU}}}\}$
& $\max\{\frac{2\Sigma_{i=1}^{N} B_{i4}\times D \times R_{i}}{C_{\text{dev}}\times N_{\text{dev}}},
\frac{D \times S \times \Sigma_{i:B_{i4}>0} R_{i}}{W_{\text{DRAM}}\times N_{\text{dev}}}\}$
& $2L_{\text{link}} + \frac{S\times\Sigma_{i=1}^{N} B_{i4}\times(D + R_{i})}{W_{\text{link}}}$ \\
\bottomrule
\end{tabular}%
}
\end{table*}

\subsection{Strategy Selection Algorithm}
\label{subsec: strategy selection algorithm}
\begin{algorithm}[t!]
\SetAlgoVlined
\footnotesize
\caption{Strategy Selection Algorithm in \xname{}}
\label{alg:choose}

\SetKwFunction{FMain}{ChooseExeStrategy}

\SetKwProg{Fn}{Function}{:}{}
\Fn{\FMain{State}}{
    strategyList = [E1, E2, E3, E4]\;
    strategyList.sort(key=CostModel)\;
    bestE = strategyList[0]\;
    Mem = MemOf(bestE)\;
    \If{$Mem < memLimit$}{
        Return bestE\;
    }
    \If{$Size(coldAdapters) > (Mem - memLimit)$}{
        Cold = FindCold(coldAdapters)\;
        bestE = bestE + Replace Cold\;
        Return bestE\;
    }
    \For{$E \in strategyList$}{
        Mem = MemOf(E)\;
        \If{$Mem > memLimit$}{
            E5 = E + Replace KV\;
            E6 = E + Replace FindCold(hotAdapter)\;
            E7 = E + Replace ServingAdapter\;
            strategyList.extend([E5, E6, E7])\;
        }
    }
    strategyList.sort(key=CostModel)\;
    Return strategyList[0]\;
}
\end{algorithm}

\xname{} selects the optimal execution strategy under the GPU memory constraint, using the cost model of~\autoref{cost model} to estimate execution time.
KV cache, LoRA adapters, and the base model compete for limited GPU memory.
We classify adapters as serving adapters used by requests in the current batch, hot adapters likely needed by future requests, and cold adapters unlikely to be used soon.
A temperature metric separates hot adapters from cold ones: each new request raises its adapter's temperature by a fixed increment, all temperatures relax toward the system average by $0.1$ of the gap per step, and \xname{} pre-caches the 50 hottest non-serving adapters.

\autoref{alg:choose} evicts less critical data when GPU memory is insufficient.
It sorts the four strategies by cost-model time (line 3) and takes the fastest that fits in GPU memory (line 6).
Otherwise it evicts cold adapters first (line 8), and if more space is still needed it also considers evicting the other GPU memory residents, namely KV cache, hot adapters, and serving adapters (lines 15-18).
Eviction itself changes performance, which complicates comparing eviction-based and non-eviction candidates, so we re-rank them all with the cost model and take the fastest (lines 19 and 20).

\subsection{Cost Model}
\label{subsec: cost model}
\label{cost model}
The cost model estimates execution time across strategies and drives the selection algorithm.
The model touches the interconnect only through its bandwidth $W_{\text{link}}$ and its round-trip latency $L_{\text{link}}$, so instantiating \xname{} on a different memory-semantic fabric such as NVLink-C2C or UALink means changing two numbers.
It covers the two major operations separately. \textbf{LoRA adapter computation} multiplies the injected weight matrix with the input vector (\autoref{subsec: lora adapter computation}) and underlies $\bm{Q}$/$\bm{K}$/$\bm{V}$ generation, projection, and FFN. \textbf{Attention computation} sits between $\bm{Q}$/$\bm{K}$/$\bm{V}$ generation and projection and follows the distributed approach of~\autoref{subsec: attention computation}.

\textbf{For LoRA computation}, the GPU and the NDP cores in the pooled-memory devices work together.
Taking $\bm{Q}$/$\bm{K}$/$\bm{V}$ generation as the representative case, $\bm{Q} = \bm{x}\bm{W} = \bm{x}(\bm{W_{\text{base}}} + \bm{A}\bm{B})$ splits into a base component $\bm{x}\bm{W_{\text{base}}}$ and a LoRA component $\bm{x}\bm{A}\bm{B}$.
The base component is GPU-only matrix multiplication that we model or profile offline, so the cost model concentrates on the LoRA component that spans the GPU and the devices.

We formulate LoRA computation with $N$ adapters of rank $R_i$, collect the request counts into a matrix whose entry $B_{ij}$ counts adapter $i$'s requests under strategy $j$, and estimate time from computation and memory-access costs.

\autoref{tab:cost-terms} lists the three cost terms of every execution strategy.
$T_{\text{GPU\_Ej}}$ is the GPU time for all requests using strategy Ej, and it absorbs E2's link fetch, which the GPU issues itself.
$T_{\text{dev\_cpt\_Ej}}$ is the NDP core computation time, which covers both the DRAM access time inside the devices and the computation time of the NDP cores.
$T_{\text{dev\_trs\_Ej}}$ is the time the devices spend returning data over the link, which covers the round-trip link latency and the pure transfer duration.
A term is zero when the strategy places no work on the corresponding resource, and a memory argument sums only over the adapters that serve a request under Ej, since an adapter loads its matrices once per step whatever its batch.
The hardware parameters are the GPU's compute throughput $C_{\text{GPU}}$ and HBM bandwidth $W_{\text{GPU}}$, the per-device NDP throughput $C_{\text{dev}}$ and DRAM bandwidth $W_{\text{DRAM}}$, and the device count $N_{\text{dev}}$. The workload parameters are the model dimension $D$, the rank $R_{i}$ of adapter $i$, the KV cache length $\ell_{k}$ of request $k$, the total batch $B = \Sigma_{i=1}^{N}\Sigma_{j=1}^{4} B_{ij}$, and the scalar $S$, the size of one value in bytes, which is 2 under FP16. Bold denotes matrices, so $S$ and $B$ are distinct from $\bm{S}$ and $\bm{B}$.
Summing across strategies gives the overall GPU time $T_{\text{LoRA\_GPU}} = \Sigma_{j=1}^{4}T_{\text{GPU\_Ej}}$ and the overall device time $T_{\text{LoRA\_dev}} = \Sigma_{j=1}^{4}T_{\text{dev\_cpt\_Ej}} + \Sigma_{j=1}^{4}T_{\text{dev\_trs\_Ej}}$.
Base-model and device computation overlap, so the total LoRA time is
\begin{equation}
T_{\text{LoRA}} = \max\{T_{\text{Base}}, T_{\text{LoRA\_dev}}\} + T_{\text{LoRA\_GPU}},\label{eq:wmm}
\end{equation}
where $T_{\text{Base}}$ is the base model GPU time, obtained by profiling.

\begin{table}[t]
\centering
\caption{Cost terms for attention computation.}
\label{tab:att-terms}
\setlength{\tabcolsep}{3pt}
\resizebox{\ifdim\width>\columnwidth \columnwidth\else \width\fi}{!}{%
\footnotesize
\begin{tabular}{@{}cl@{}}
\toprule
$T_{\text{ATT\_GPU}}$ & $P_{\text{KV}}\times \max\{\frac{4D \times \Sigma_{k=1}^{B} \ell_{k}}{C_{\text{GPU}}},\frac{2D  \times \Sigma_{k=1}^{B} \ell_{k} \times S}{W_{\text{GPU}}}\}$ \\[2pt]
$T_{\text{ATT\_dev}}$ &
\begin{tabular}[t]{@{}l@{}}
$T_{\text{dev\_cpt}} + T_{\text{dev\_trans}}$ \\
$=(1 - P_{\text{KV}}) \times\max\{\frac{4D \times  \Sigma_{k=1}^{B} \ell_{k}}{C_{\text{dev}}\times N_{\text{dev}}},\frac{2D \times S\times\Sigma_{k=1}^{B} \ell_{k} }{W_{\text{DRAM}}\times N_{\text{dev}}}\}$ \\
$\quad+ 2L_{\text{link}} + \frac{2D\times B \times S}{W_{\text{link}}}$
\end{tabular} \\[2pt]
$T_{\text{ATT}}$ & $\max\{T_{\text{ATT\_GPU}}, T_{\text{ATT\_dev}}\}$ \\
\bottomrule
\end{tabular}%
}
\end{table}

\textbf{For attention computation}, the GPU and each device compute the entire $(softmax(\bm{Q}\times \bm{K}^{T})\times \bm{V})$ over disjoint input partitions, and the GPU aggregates the results.
$P_{\text{KV}}$ is the fraction of KV cache duplicated in GPU memory, so each device handles $\frac{1-P_{\text{KV}}}{N_{\text{dev}}}$ of the KV cache-related computation.
\autoref{tab:att-terms} lists the cost terms, where $T_{\text{ATT\_GPU}}$ and $T_{\text{ATT\_dev}}$ both account for memory access and computation and the total attention time $T_{\text{ATT}}$ takes their maximum.

Aggregating the LoRA and attention costs across all layers yields the total execution time that the strategy selection algorithm uses to rank every candidate decision and pick the best one for each request.

\begin{figure*}[t]
    \centering
    \includegraphics[width=\textwidth]{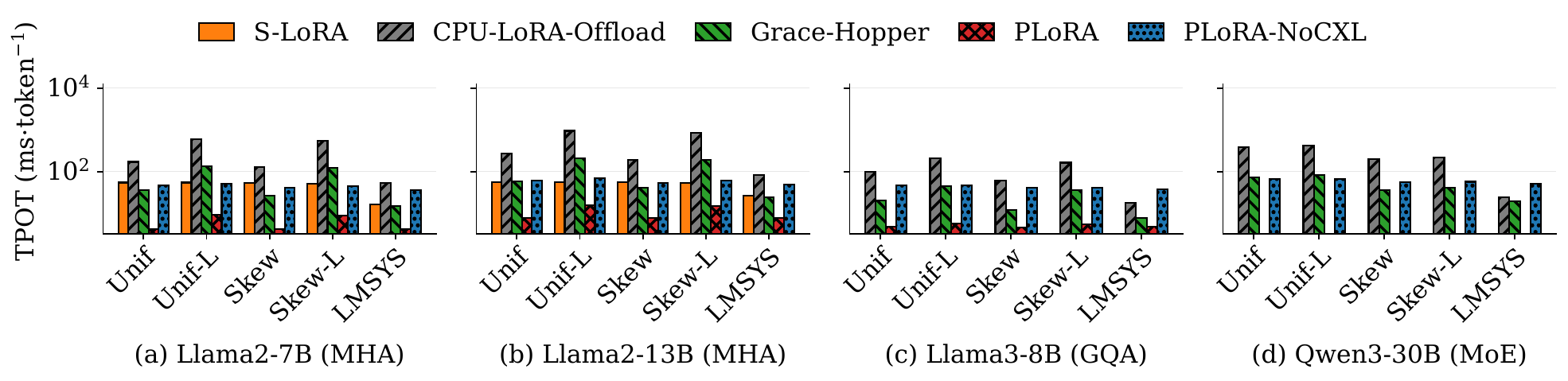}
    \caption{Decode latency (TPOT, lower is better) on H100. S-LoRA is a real-machine measurement and appears only for the two models we could run.}
    \label{fig:rebuttal-decode-tpot}
    \vspace{-8pt}
\end{figure*}

\section{Evaluation}
\subsection{Experiment Setup}
\label{subsec: exp-setup}
\noindent\textbf{Methodology: } We implement the \xname{} GPU memory manager, a strategy selection algorithm plus a cost model, in Python (1200 lines). Lacking an established CXL simulator, we built an event-driven multi-device simulator in C (about 5,700 lines) extending SSDSim~\cite{ssdsim1, ssdsim2} with redesigned request and hardware-module state machines. We calibrate on real hardware: GPU utilization across batch sizes gives compute time, and memory access patterns on Samsung CMM-D, an NDP-less CXL device, give timing parameters.

\textbf{Hardware Settings: }
We use one NVIDIA {H100} GPU {(80 GB)} with {3350 GB/s} bandwidth and four \xname{} memory devices, each attached to the CXL switch over CXL 3.1 at {128 GB/s}.
Device configurations appear in~\autoref{tab:system-config-updated} and never exceed the values Samsung reports~\cite{park2024lpddr_cxl}.

\textbf{Model Settings: }
We evaluate Llama2-7B and Llama2-13B, both using multi-head attention (MHA)\cite{llama2}. For generality we add Llama3-8B, which uses group query attention (GQA)\cite{llama3}, and Qwen3-30B, which combines GQA with a mixture-of-experts FFN.
Each LoRA adapter takes a dimension randomly chosen from [8, 16, 32, 64, 128], and we add adapters to all weight matrices, including Q/K/V, projection, and FFN.

\textbf{Baselines: } We compare against five systems. S-LoRA~\cite{sheng2023s-slora} is the published multi-LoRA serving system closest to our setting. Pure CXL uses the same devices as plain memory expanders with no NDP, which isolates what near-data execution contributes. CPU-LoRA-Offload stages adapters through host memory over PCIe. A Grace-Hopper configuration tests whether a faster host link alone suffices, at 450~GB/s. PLoRA-NoCXL keeps our hardware but removes accelerator load/store access, so it must launch a kernel per adapter. We omit attention accelerators like AttAcc~\cite{park2024attacc}, which help attention but leave adapter capacity, the binding constraint, untouched.

\begin{table}[tb]
\centering
\caption{System and cost-model configuration details.}
\label{tab:system-config-updated}
\setlength{\tabcolsep}{4pt}
\footnotesize
\begin{tabular}{@{}ll@{}}
\toprule
Component & Configuration \\ \midrule
GPU & 1 $\times$ NVIDIA H100, 80 GB, 3350 GB/s \\
GPU compute & 989 TFLOPS FP16 \\
Mem device & 4 \xname{} devices, 512 GB each \\
Device DRAM & 8-package LPDDR5X DIMM, 1.1 TB/s per device \\
Per package & 64 GB, 136 GB/s \\
CXL link & PCIe 6.0 and CXL 3.1, 128 GB/s, 200 ns latency \\
NDP core & 2 TFLOPS FP16 per device, 8 TFLOPS total \\
NDP buffer & 3 MB on-chip \\
\bottomrule
\end{tabular}
\vspace{-6pt}
\end{table}

\textbf{Workloads: } \autoref{7-tab: dataset-config} lists our real and synthetic workloads.
The real one is LMSYS Chatbot Arena, a multi-model serving dataset also used by prior multi-LoRA work~\cite{sheng2023s-slora}, where we treat each model as a LoRA variant. Of the synthetic datasets, uniform ones spread requests evenly across adapters, skewed ones send 80\% of requests to 50 adapters, and `long' ones use larger sequence lengths.

\textbf{Configurations: } For the ablation study, three capital letters denote our three techniques: \textbf{N} (NDP capabilities), \textbf{S} (strategy selection algorithm), and \textbf{D} (distributed attention computation). This notation yields:
\circlednumber{black}{1} \textbf{N(E3), N(E4)}: NDP cores only, with E3 or E4 applied uniformly to all requests.
\circlednumber{black}{2} \textbf{NS}: NDP cores plus strategy selection, no distributed attention.
\circlednumber{black}{3} \textbf{ND(E3), ND(E4)}: NDP cores plus distributed attention, no strategy selection, with E3 or E4 applied uniformly.
\circlednumber{black}{4} \textbf{NSD (Ours)}: complete \xname{}.

\begin{table}[tb]
\centering
\caption{Workload Configuration Details}
\footnotesize
\setlength{\tabcolsep}{3.5pt}
\begin{tabular}{@{}ccccc@{}}
\toprule
 Workload           & Req Distr. & LoRA Num & Input Len        & Output Len        \\ \midrule
LMSYS        & skewed           & 25           & {[}2, 430{]}     & {[}2, 430{]}     \\
Uniform      & uniform          & 1000         & {[}100, 1024{]}  & {[}100, 1024{]}  \\
Uniform-long & uniform          & 1000         & {[}2048, 4096{]} & {[}2048, 4096{]} \\
Skewed       & skewed           & 1000         & {[}100, 1024{]}  & {[}100, 1024{]}  \\
Skewed-long  & skewed           & 1000         & {[}2048, 4096{]} & {[}2048, 4096{]} \\ \bottomrule
\end{tabular}
\label{7-tab: dataset-config}
\vspace{-10pt}
\end{table}

\subsection{Decode Performance}
\xname{} attains the lowest TPOT (time per output token) on H100 for every workload and all four base models in~\autoref{fig:rebuttal-decode-tpot}, and it runs 3.7$\times$ to 177$\times$ below CPU-LoRA-Offload. S-LoRA is measured on a real H100, so we report it only for the two models we could run, where \xname{} averages 6.6$\times$ below it and reaches 13.1$\times$. Grace-Hopper's 450~GB/s C2C link competes only at short context, coming within 1.6$\times$ to 3.6$\times$ of \xname{} on the Llama models under LMSYS, but degrades sharply on long context, 13.6$\times$ slower on Llama2-13B Uniform-long. PLoRA-NoCXL runs 4$\times$ to 11$\times$ slower than \xname{} on the dense models and up to 31$\times$ slower on the MoE model, isolating the benefit of CXL itself. The MoE model preserves this ordering because its experts are memory-bound.
\xname{} also holds across attention mechanisms and FFN structures: it keeps the lowest TPOT on Llama3-8B and Qwen3-30B, and at long context Llama3-8B beats Llama2-7B because GQA shrinks the KV cache.

\begin{table}[tb]
\centering
\caption{Prefill TTFT (s) of \xname{} on H100. Skewed workloads track uniform within 4\%.}
\label{tab:prefill-ttft}
\setlength{\tabcolsep}{2.5pt}
\footnotesize
\begin{tabular}{@{}lcccccc@{}}
\toprule
 & \multicolumn{3}{c}{Unif} & \multicolumn{3}{c}{Unif-L} \\
\cmidrule(lr){2-4} \cmidrule(l){5-7}
Model & B=256 & B=512 & B=1024 & B=256 & B=512 & B=1024 \\ \midrule
Llama2-7B, Llama3-8B  & 2.54 & 5.03 & 10.26 & 15.33 & 30.87 & 62.41 \\
Llama2-13B & 4.93 & 9.77 & 19.93 & 29.25 & 58.90 & 119.07 \\
Qwen3-30B  & 0.98 & 1.94 & 3.95 & 6.40 & 12.91 & 26.11 \\
\bottomrule
\end{tabular}
\vspace{-6pt}
\end{table}

\setcounter{topnumber}{1}
\subsection{Prefill Performance}
\autoref{tab:prefill-ttft} reports prefill latency (TTFT) for Llama2-7B and Llama2-13B (MHA), Llama3-8B (GQA), and Qwen3-30B (MoE) at batch sizes 256, 512, and 1024. TTFT stays within a few seconds at batch 256 on short sequences and reaches about 120\,s only at batch 1024 on the 13B long-sequence workloads, outside the target serving regime. Llama2-7B and Llama3-8B share a row because our prefill model resolves the forward pass at hidden size and layer count, which the two share, so it does not separate Llama3-8B's larger FFN.

\subsection{Ablation Study}

\begin{figure}[t]
    \centering
    \includegraphics[width=\columnwidth]{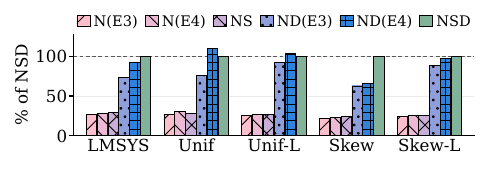}
    \caption{Ablation on Llama2-7B, bars as a percentage of full \xname{} (NSD).}
    \label{fig:exp-ablation}
    \vspace{-6pt}
\end{figure}

\textbf{Benefit of distributed attention (versus NS): }
\xname{} averages 3.8$\times$ over NS in~\autoref{fig:exp-ablation}, from 3.5$\times$ on LMSYS to 4.1$\times$ on Skewed. NS leaves attention entirely on the GPU, so it drags the KV cache across the link on every workload.

\textbf{Benefit of strategy selection (versus ND(E3) and ND(E4)): }
\xname{} beats ND(E3) by 30\% on average and by 59\% on Skewed, where a fixed strategy serves the crowded and the idle adapters alike. It leads ND(E4) by 10\% on average, from 52\% ahead on Skewed to 10\% behind on Uniform. ND(E4) is idealized on the 1000-adapter workloads because E4 there caches one LoRA matrix per layer and per injected weight, 91\,GB at their mean rank, which no H100 holds.

\textbf{Benefit of both techniques (versus N(E3) and N(E4)): }
\xname{} averages 4.0$\times$ over N(E3) and 3.8$\times$ over N(E4), rising to 4.7$\times$ and 4.3$\times$ on Skewed.

\subsection{Breakdown Study}
\label{ssec: breakdown study}
\begin{figure}[t]
    \centering
    \subfloat[]{\includegraphics[width=0.49\linewidth]{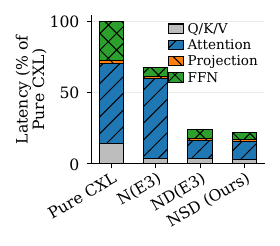}\label{fig:exp-breakdown}}
    \hfill
    \subfloat[]{\includegraphics[width=0.49\linewidth]{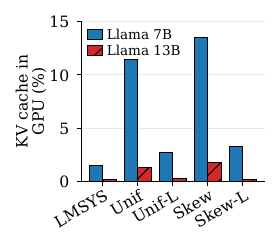}\label{fig:exp_gpu_mem}}
    \caption{(a) Latency breakdown for one 7B layer, as relative shares. (b) KV cache fraction kept in GPU memory.}
    \vspace{-8pt}
\end{figure}

\textbf{Latency Breakdown:} In~\autoref{fig:exp-breakdown}, N(E3) moves LoRA operations to the NDP cores and cuts one layer's latency by about a third over Pure CXL, ND(E3) adds collaborative computation, which alone divides the attention component by 4.5$\times$ and removes a further 65\%, and NSD (Ours) adds strategy selection and removes another 8\%.

\textbf{KV Cache Breakdown:} \xname{} duplicates 0.2\%--13.5\% of the KV cache in GPU memory in~\autoref{fig:exp_gpu_mem}, relieving the CXL devices. The 13B model stays under 2\%, since its larger weights leave less GPU memory for adapters and KV cache.

\subsection{Sensitivity Study}

\begin{figure}[t]
    \centering
    \includegraphics[width=\columnwidth]{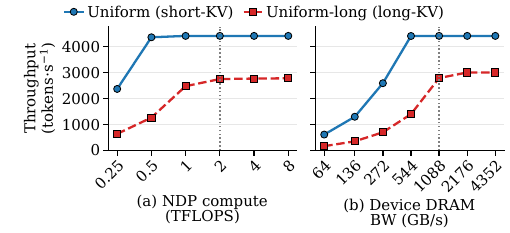}
    \caption{Sensitivity of \xname{} decode throughput to the two device-side parameters that bind, (a) NDP compute and (b) device DRAM bandwidth.}
    \label{fig:rebuttal-sensitivity}
    \vspace{-8pt}
\end{figure}

\autoref{fig:rebuttal-sensitivity} sweeps the two binding device-side parameters under a short-KV (Uniform) and a long-KV (Uniform-long) workload. \xname{} saturates once NDP compute exceeds about 1--2\,TFLOPS~(a), validating our conservative 2\,TFLOPS per device, and scales with device DRAM bandwidth~(b) to a plateau beyond which the kernel turns compute-bound.
Throughput is also flat in NDP buffer size beyond two on-chip buffers, execution being device-DRAM bound. Link latency and bandwidth get their own treatment in~\autoref{sec: link-generality}.
On H100 at batch 32, GPU compute takes 0.45/0.52\,ms (Llama2-7B/Llama3-8B) against 4.18/4.78\,ms to read weights from HBM, so FLOPs stop helping once the NDP core keeps up with device DRAM.

KV cache compression does not weaken the case for \xname{}. On Llama2-13B under Uniform-long, \xname{} remains the fastest at every precision, and its margin over PLoRA-NoCXL, the strongest baseline at all three, widens from 4.4$\times$ under an FP16 KV cache to 6.3$\times$ under INT8 and 7.8$\times$ under INT4, because a smaller KV cache shifts pressure to adapter handling, where near-data execution helps most.

\subsection{Comparison with Multi-GPU Baselines}
\label{exp-multi-gpu}
Throughput of multi-GPU S-LoRA on Llama2-7B improves with GPU count, yet \xname{} still outperforms 4-GPU S-LoRA by $2.23\times$.
As an exploratory estimate, assuming \$30,000 per H100 and, lacking a commercial CXL+NDP part, \$3,000 per device for 512\,GB LPDDR5X plus a top-tier FPGA, a conservative bound since the NDP logic adds only 3.4\% area and the estimate excludes the CXL switch, host platform, and cooling, one accelerator plus four devices costs \$42,000 against \$120,000 for four accelerators, which puts \xname{} $6.4\times$ ahead of 4-GPU S-LoRA on throughput per dollar. S-LoRA's throughput grows sublinearly in GPU count while its cost grows linearly, so the gap widens with every GPU added.

\begin{figure}[t]
    \centering
    \includegraphics[width=\columnwidth]{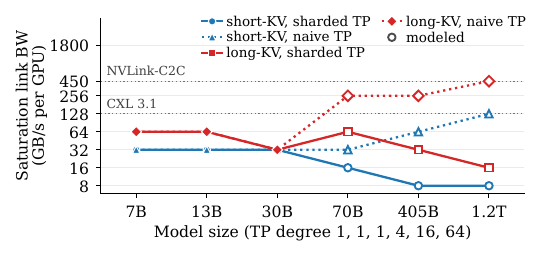}
    \caption{Saturation link bandwidth against model scale. Filled markers are simulator ladders, open markers the calibrated tensor-parallel model.}
    \label{fig:bw-scaling}
\vspace{-8pt}
\end{figure}

\subsection{Area and Power Overhead}
We implement \xname's NDP core in Verilog HDL and synthesize it with Design Compiler in TSMC 45nm, with 3MB of SRAM for input, DRAM, and output buffers plus 1024 FP16 multiply-add units in the PEs delivering 2048 GFLOPS at 1GHz. Buffers dominate at 44.22\,mm$^2$ and 1.91\,W, and the PEs add 0.80\,mm$^2$ and 1.17\,W, totaling 45.02\,mm$^2$ and 3.08\,W. Against an original DIMM above 1344\,mm$^2$ and 40\,W, area and power overhead stays below 3.4\% and 7.7\%.

\section{Trading Link Bandwidth for Near-Data Compute}
\label{sec: link-generality}

Every result so far assumed a 128\,GB/s CXL 3.1 link. We now sweep link
bandwidth over nine points from 8 to 1800\,GB/s, spanning the PCIe 5.0, CXL 3.1,
NVLink-C2C, NVLink 5 and NVLink 6 classes, holding all else fixed. Every system moves its link
constant with the x-axis except \xname{}, conservatively: its
selection policy stays tuned for 128\,GB/s.

\begin{figure*}[t]
    \centering
    \includegraphics[width=\textwidth]{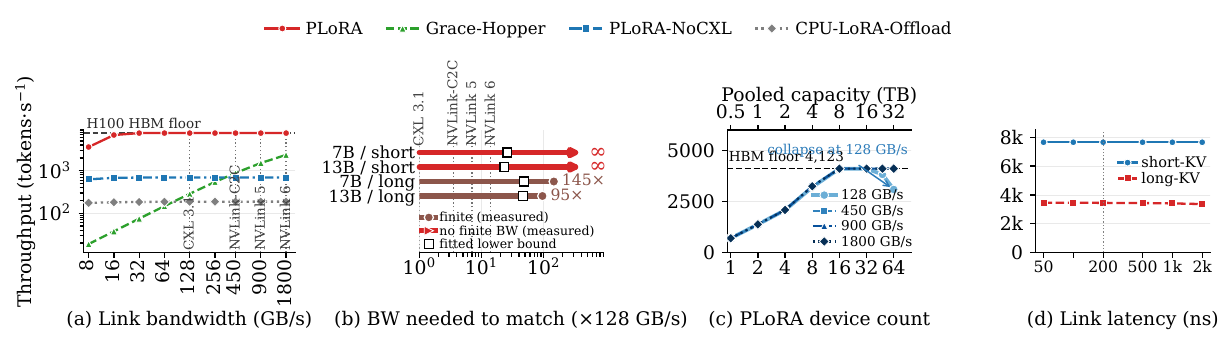}
    \caption{\xname{} across the interconnect ladder: (a) throughput against link bandwidth (Llama2-7B, Uniform; rules mark per-direction fabric rates), (b) bandwidth a no-NDP system needs to match \xname{}, (c) throughput against device count (Llama2-13B, Uniform-long), and (d) per-layer one-way link latency.}
    \label{fig:link-ladder}
\end{figure*}

\noindent\textbf{Finding 1: \xname{} stops responding to link bandwidth by 32
to 64\,GB/s, and on short contexts its plateau is the accelerator's own HBM floor.}
\xname{} saturates at 32\,GB/s on short contexts and by 64\,GB/s on long, for
both models, one of which \autoref{fig:link-ladder}(a) plots. Raising the link a further 14$\times$, from the
128\,GB/s we assume to NVLink 6's 1800\,GB/s, moves throughput by between
$+0.00\%$ and $+0.39\%$.
The plateau is the more informative number: on short contexts a decode step settles at
4.18\,ms on Llama2-7B and 7.76\,ms on Llama2-13B, exactly the time to stream
14\,GB and 26\,GB of weights from HBM at 3350\,GB/s, which is 7,657 and
4,123\,tokens/s at the batch of 32 the sweep uses. Long contexts plateau
lower, at 3,461 and 2,092\,tokens/s, where the KV cache sets the step time. \xname{} does not merely reduce link traffic, it retires the link as a
bottleneck and leaves the system bound by the accelerator's own memory, which is
the bound a GPU with unlimited local capacity would also have. PLoRA-NoCXL saturates by 32\,GB/s too, at 468 to 690\,tokens/s, because
per-adapter kernel launches bound it rather than bandwidth, leaving \xname{}
4.0$\times$ to 11.1$\times$ faster at every point on the ladder, and CPU-LoRA-Offload is
CPU-bound below 188\,tokens/s. Grace-Hopper alone tracks the link, rising
with bandwidth at a slope that decays.

\setcounter{topnumber}{2}
\noindent\textbf{Finding 2: buying the same result with bandwidth alone is not
practical.} If \xname{} does not need bandwidth, how much would it take to do
without \xname{}? We sweep Grace-Hopper, the fastest link-only baseline, whose
throughput bends sublinear before the ladder tops out, so we measure the
crossing by bisection rather than extrapolation.
\autoref{fig:link-ladder}(b) reports the result. On the long-context workloads it
needs 12.2\,TB/s for Llama2-13B and
18.6\,TB/s for Llama2-7B to match \xname{} on a 128\,GB/s link, which is
95$\times$ to 145$\times$ a CXL 3.1 link and 7$\times$ to 10$\times$ beyond
NVLink 6. On the short-context workloads it never matches \xname{} at all: its
throughput saturates 1.03$\times$ to 1.46$\times$ short at any bandwidth,
limited by per-kernel launch overhead rather than the link.
In cost terms the near-data engine that buys two orders of magnitude of
effective interconnect is 45\,mm$^2$, under 3.4\% of the device beside it,
while the bandwidth it replaces is the least divisible resource in the
machine.

\noindent\textbf{Finding 3: surplus link bandwidth becomes capacity, not speed.}
What a wider link buys is neither peak performance nor a better operating
point: across 128 to 1800\,GB/s the peak stays at the HBM floor and the knee
stays at 4 devices on short KV and 16 on long. Our four-device configuration
sits at the short-context knee and short of the long-context one, so the
long-context results above understate a larger pool.
What the wider link removes is the collapse beyond the knee.
On a 128\,GB/s link, \autoref{fig:link-ladder}(c) shows throughput holding to 32
devices and then falling, by 25\% at 64 and by 38\% on short KV, because
each device returns one partial result for the GPU to aggregate, so aggregation
traffic grows with the pool while the link does not. At 450\,GB/s the collapse
disappears and full throughput persists to 64 devices, our simulator's largest
pool. The usable pool therefore at least doubles, from 32 devices and 16\,TB
to 64 and 32\,TB at 512\,GB per device.
The gain saturates at 450\,GB/s, so NVLink-C2C bandwidth captures it,
and the width that buys a bandwidth-bound system speed buys \xname{}
adapters and context, the currency the workload lacks.

\noindent\textbf{Finding 4: under direct memory access, link latency is second order.}
\autoref{fig:link-ladder}(d) sweeps the one-way link latency from 50 to 2000\,ns, charged per layer since decode is layer-sequential, two round trips per layer, one LoRA wave and one attention return.
Short-KV throughput never moves, the charge hides under the base-model step, and the CXL-bound long-KV workload gives up 0.3\% at our assumed 200\,ns and 2.6\% at 2000\,ns.
This does not contradict \autoref{subsec: why is CXL better?}: a memory-semantic link pays these nanoseconds from inside the running kernel, where in-flight loads overlap them, while the PCIe+CPU path pays microsecond software latencies, a kernel stop, a host-run copy, and a relaunch on every exchange, and it forfeits near-data reduction.
CXL-class fabrics win on the access model, not on wire latency, so \xname{} tolerates any switch-class latency.

\noindent\textbf{Finding 5: the verdict survives model scale when the parallelism cooperates.}
\autoref{fig:bw-scaling} asks what happens beyond one GPU: it sweeps the saturation bandwidth, the smallest ladder point within 1\% of the plateau, from 7B to a 1.2T dense model.
The three single-GPU models are full simulator ladders, and the MoE model saturates at 32\,GB/s even on long contexts, below the dense models' 64, because its step is floored by streaming its 6\,GB of active experts.
The three larger models come from a simulator-calibrated tensor-parallel model: weights, heads, and the HBM floor shard across TP-4 to TP-64 GPUs, each GPU keeps its own pool, the all-reduce rides the inter-GPU fabric, and the link coefficients are fitted from the simulated ladders, reproducing them within 1\% on the reference model.
Sharding decides the outcome: if each GPU replicates the full activation into its pool on every adapter wave, the naive port of tensor parallelism, per-GPU demand climbs with scale and reaches 450\,GB/s at 1.2T on long contexts, because the HBM floor thins as $1/N$ while the wave input does not.
Sharding the adapter wave with the model instead, a row-parallel $\bm{A}$, an $R$-wide all-reduce, then a column-parallel $\bm{B}$, sends demand falling with scale, down to 8 or 16\,GB/s per GPU at 1.2T, because pooled traffic then thins as $1/N$ too while the deeper floor hides it.
The rule for large deployments is one sentence: partition the adapter waves with the tensor parallelism, and pooled memory needs less bandwidth per GPU the bigger the model.

\noindent\textbf{Scope of the claim.} Four caveats. The sweep varies the link
with the device side fixed, which isolates the link's contribution but
understates a real fabric generation, whose devices would also carry more
internal bandwidth and NDP throughput, the resources
\autoref{fig:rebuttal-sensitivity} identifies as binding. The doubled pool is a
lower bound, the 450\,GB/s curve still flat where the simulator runs out of
devices. No NVLink-attached part is yet built for memory
expansion~\cite{werner2025nvlinkc2c}, so the NVLink and UALink rows of
\autoref{tab:fabrics} say where the design transfers rather than what we
measured. And TP-64 abstracts a pod that would mix tensor with pipeline
parallelism, which only lowers per-GPU pooled traffic.

\section{Related Work}
\textbf{LLM and multi-LoRA serving.} Serving frameworks~\cite{yu2022orca, kwon2023vllm, zheng2023sglang} and techniques like speculative decoding~\cite{specinfer} and offloading~\cite{sheng2023flexgen} target LLM inference. For multi-LoRA, S-LoRA~\cite{sheng2023s-slora}, PUNICA~\cite{chen2024punica}, dLoRA~\cite{wu2024dlora}, and CaraServe~\cite{li_caraserve_2024} improve adapter execution, GPU usage, merging, and CPU offloading, but all keep capacity behind a host-mediated link.

\textbf{CXL.} Work on CXL spans device characterization~\cite{sun2023demystifying_type1, ji2024demystifying_type2}, OS tiering support~\cite{maruf2023tpp, zhou2024neomem_tiering}, disaggregation~\cite{li2022pondcxl} and page management~\cite{huo2024pifs}, DRAM~\cite{park2024lpddr_cxl} and Flash~\cite{yang2023ssd_cxl} devices, and acceleration for DLRM training~\cite{recxl}, distributed learning~\cite{deepmemorydl}, genomics~\cite{huangfu2022beacon}, ANN search~\cite{jang2023cxl_anns}, and analytics~\cite{ryu2023system}.

\textbf{Near data processing.} NDP on DRAM splits into PIM, which exploits bank parallelism but needs process changes~\cite{liu2025makellm_pim1, pim2, lee2021pim3, pim4, pim5, pim6}, and PNM, which adds rank-level logic beside the DIMM and preserves the DRAM design~\cite{ke2021pnm1, pnm2, pnm3, ham2024lowoverhead_pnm4, kwon2019tensordimm_pnm5, pnm6}. \xname{} uses PNM, and prior CXL-PNM platforms, the device we follow~\cite{park2024lpddr_cxl} and Marvell's Structera A~\cite{marvell_structera}, supply a general offload substrate. Our delta is the decision layer above it: which operator to place near data, for which adapter, at which batch size, under a GPU memory budget.

\textbf{Memory-semantic fabrics.} Prior work measures what CXL and NVLink-class fabrics deliver~\cite{li2020gpuinterconnect, werner2025nvlinkc2c}, and \autoref{subs:CXL} surveys their convergence on load/store pooling~\cite{nvidia_fam_patent1, nvidia_fam_patent2, cuda_egm, nvlink_fusion, ualink}. To our knowledge no serving system targets their shared properties rather than one of them, or reports how a near-data design behaves across their bandwidth range.

\section{Conclusion}
Multi-LoRA serving asks a GPU for orders of magnitude more capacity per FLOP than it supplies, and moving that capacity behind a narrow link relocates the problem.
\xname{} spends near-data compute in place of link bandwidth, an exchange this workload makes cheap because its memory-hungry operators are also its most reducible.
Once the reduction happens at the memory the link leaves the critical path, so CXL is this design's openly specified carrier rather than its premise, and a faster fabric buys pooled capacity, not speed, a verdict that strengthens as models scale into tensor-parallel pods.
That inversion, from bandwidth-limited to capacity-limited, says where memory-semantic fabrics should spend their silicon.


\bibliographystyle{IEEEtranS}
\bibliography{main}


\end{document}